\documentclass[a4paper,10pt]{article}

\usepackage[spacing]{microtype}
\usepackage[T1]{fontenc}
\usepackage{amsfonts,textcomp} 
\usepackage{amssymb,amsmath,amsthm}
\usepackage{textcase}
\usepackage{xcolor,hyperref}
\usepackage[round]{natbib}
\usepackage{units}
\usepackage{authblk}
\usepackage{prolog,abbrvs,refutils,figutils,maths,nicefrac}

\def\nf(#1,#2){\nicefrac{#1}{#2}}

\definecolor{DarkRed}{rgb}{0.4,0.1,0}
\definecolor{MidRed}{rgb}{0.6,0,0}
\hypersetup{colorlinks,citecolor=black,urlcolor=MidRed,linkcolor=MidRed}

\usepackage{graphicx}

\DeclareMathAlphabet{\mathcal}{OMS}{cmsy}{m}{n}

\def\TheAuthors{Samer Abdallah (samer@jukedeck.com)}
\def\TheAcknowledgments{I would like to thank Jan Burse for discussions on the SWI Prolog
mailing list about using Prolog for differentiation and for prompting me to
investigate the problem of computing multiple derivatives (for the purpose of finding
Taylor series coefficients), which led to the second, improved implementation presented here).}

\def\TheAuthors{Samer Abdallah (\texttt{samer@jukedeck.com})}
\def\TheInstitution{Jukedeck Ltd.}

\def\TheTitle{Automatic Differentiation using Constraint Handling Rules in Prolog}

\def\TheAbstract{Automatic differentiation is a technique which allows a programmer
to define a numerical computation via compositions of a broad range of
numeric and computational primitives and have the underlying system support
the computation of partial derivatives of the result with respect
to any of its inputs, without making any finite difference approximations, and without
manipulating large symbolic expressions representing the computation. This note describes 
a novel approach to reverse mode automatic differentiation using constraint logic programmming,
specifically, the constraint handling rules (CHR) library of SWI Prolog,
resulting in a very small (50 lines of code) implementation. 
When applied to a differentiation-based implementation of the inside-outside
algorithm for parameter learning in probabilistic grammars, the CHR based implementations
outperformed two well-known frameworks for optimising differentiable functions, Theano and TensorFlow, 
by a large margin.}

\title{\TheTitle}
\author{\TheAuthors}
\affil{\TheInstitution}
\date{\textordinal{31}{st} May, 2017}

\begin{document}
	\maketitle
	\begin{abstract}\TheAbstract\end{abstract}
	\section{Introduction}
\seclab{intro}
Automatic differentiation (AD, also known as \emph{algorithmic differentiation}
or \emph{computational differentiation}) is a technique that enables the computation of
of the partial derivatives of numerical computations with respect to
any of their inputs. In many cases, this is to be preferred to
a process of manual differentiation and coding, which, although mechanical
and not very difficult, can quite become quite laborious and error-prone 
when the expressions being differentiated are large and complex.

Other solutions to this problem include approximations based on finite
differences and symbolic differentiation. In comparison, AD
does not involve the approximations of finite difference methods (other
than the approximation inherent in the use of fixed-width floating-point numbers).
The distinction between AD and symbolic differentiation
(of the kind one might do with computer algebra systems such as Mathematica)
is more or less clear-cut depending on which variant of AD is being used: 
so-called \emph{forward mode} AD can be implemented by augmenting
the numerical data type to include derivatives (so-called \emph{dual numbers})
and overloading the primitive numerical operators and functions to handle the new 
data type, without further modification of the code and without requiring any explicit representation 
of the computation as a symbolic expression (other than the code itself).
\emph{Reverse mode} AD, however, requires a more
explicit representation of the computation as a graph, which must be built and
then traversed both forwards and backwards to get numerical results for the function
and its derivative. A symbolic representation of the expression and its derivatives 
is, arguably, implicit in the graph, though not, perhaps, in the sort of form  
one might manipulate in a computer algebra system. 

AD is large, well-developed field; for a textbook introduction, see, \eg,
\cite{GriewankWalther2008}, or, for a more recent review targeted at the
machine learning community, \cite{BaydinPearlmutterRadul2015}. 

In recent years, symbolic and automatic differentiation have become an important component
of many machine learning frameworks such as Torch \citep{CollobertKavukcuogluFarabet2011}, 
Theano \citep{2016arXiv160502688short} 
and TensorFlow \citep{MartinAbadiAshishAgarwalPaulBarham2015}, because the 
minimisation of some differentiable loss function is at the heart of many 
machine learning models, and these loss functions are increasingly complex,
arising from the composition of many modular parts. These frameworks tend to use
reverse mode AD, since the functions to be differentiated are usually scalar-valued 
with many (hundreds or thousands) parameters, and reverse mode is more efficient than
forward mode in this regime. Indeed, the well-known \emph{back-propagation} algorithm
for training neural networks \citep{RumelhartHintonMcClelland1986} is nothing more than
reverse mode AD confined to a limited set of vector-to-vector and vector-to-scalar operators.

A lesser known observation is that the \emph{outside algorithm} \citep{Goodman1998},
used in conjunction with the \emph{inside algorithm} to fit the parameters of a probabilistic 
context free grammar to a given corpus, is also essentially reverse mode differentiation
of the top-level probability produced by the inside algorithm with respect to the 
parameters (a collection of discrete probability distributions) of the grammar.
To my knowledge, this was first noticed by \cite{SatoKameya2001}, who, in generalising 
the inside-outside algorithm to their probabilistic programming
system PRISM, expressed the outside probabilities as the partial derivatives
of the topmost inside probability with respect to the parameters of the model.
In fact, one can go slightly further than they did: computing the derivatives of the
\emph{logarithm} of the topmost inside probability with respect to the \emph{logarithms} of
the model parameters yields directly the sufficient statistics for updating the parameters
in an expectation maximisation algorithm.

In this note, I describe a novel approach to reverse mode AD based on constraint logic programming,
specifically, an implementation using constraint handling rules (CHR) in SWI Prolog 
\citep{Fruhwirth1998,WielemakerSchrijversTriska2012}.
Some familiarity with Prolog and CHR is assumed---see \cite{SneyersVan-WeertSchrijvers2010} for an
introduction and survey.
Although not intended to compete in terms of performance and breadth of scope with existing
AD systems, it is extremely succinct and could form the basis of a more practically
useful system in future, for example, by marrying the high-level CHR-based front-end 
with a high-performance multi-core or GPU backend for the low-level numerical 
operations.

I will describe two implementations of the idea, in the order they were developed,
in order to describe a limitation of the former method which is solved in the 
latter. As both implementations amount to less than 100 lines of CHR/Prolog in total,
the complete code is presented below, with numbered lines and framed between horizontal rules.
Interactions with the SWI Prolog are displayed in a fixed-width typeface with a grey
bar on the left.

\section{First attempt: forward constraint propagation}
\seclab{autodiff1}

The module preamble declares the exported predicates (which are all CHR constraints),
loads the CHR module, and declares the CHR constraints with their modes (the symbols "-"
and "+" indicate that the corresponding argument must be an unbound variable or a
ground term respectively, while "?" means the argument can be either):
\begin{prolog-framed}[name=adone]
	:- module(autodiff1, [mul/3, add/3, pow/3, exp/2, log/2, deriv/3, go/0]).
	:- use_module(library(chr)).
	:- chr_constraint add(?,?,-), mul(?,?,-), log(-,-), exp(-,-), pow(+,-,-),
										deriv(?,-,?), agg(?,-), acc(?,-), acc(-), go.
\end{prolog-framed}
The interface constraints "add/3", "mul/3", "pow/2", "exp/2" and "log/2" provide
the arithmetic primitives, and are intended to be used with Prolog variables to
define the desired computation, which can be thought of as a \emph{hypergraph} with
a hyperedge for each constraint\footnote{%
In the sequel, for brevity, I will simply refer to these as ``graphs'', and in
the visualisations which follow, the hyperedges will be rendered as nodes in boldface text, 
with each kind of hyperedge having a particular collection of inputs and outputs
corresponding to the arguments of the corresponding CHR constraint.},
for example, the hyperedge "mul(X,Y,Z)" connects
the nodes "X", "Y" and "Z" and means that "Z" is the product of "X" and "Y".
Then, "deriv/3" is used to request the partial
derivative of one variable with respect to another and "go/0" is used to trigger
a process in which arithmetic constraints between the requested derivatives and
other variables in the graph are established.

This code is already runnable, and will allow the building of a passive computation
graph, as the following interaction in SWI Prolog demonstrates (assuming the code has 
been saved in a file called \texttt{autodiff1.pl}):
\begin{prolog-barred}
	?- use_module(autodiff1).
	?- mul(A,B,C), add(1,C,D), log(D,E).
	add(1, C, D),
	mul(A, B, C),
	log(D, E).
\end{prolog-barred}
The three lines printed after the query display the contents of the constraint store,
which we can visualise as a graph:
\begin{center}
\includegraphics{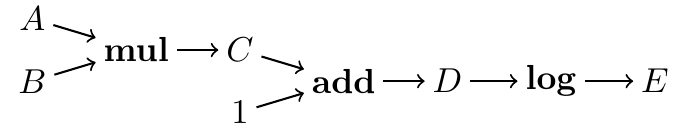}
\end{center}
This example also illustrates how "add/3" (and "mul/3") can accept ground arguments as well
as variables: these represent constants in the computation graph, and support the use of
standard Prolog high-order programming constructs in a natural way to compose arithmetic primitives
into more complex computations, for example, a list of variables "Xs" can be summed using
"foldl(add, Xs, 0, Sum)".

Next, a few CHR simplification rules ($\Leftrightarrow$) rules handle algebraic axioms such as
$x + 0 = 0$ and $x^1 = x$, propagation rules ($\Rightarrow$) use "delay/2" to set up 
delayed Prolog goals for evaluating expressions numerically once the operands are
grounded, and simpagation rules ("_ \ _ <=> _") remove duplicate constraints from the store.
\begin{prolog-framed}[name=adone,firstnumber=5]
	mul(1,X,Y) <=> Y=X.
	mul(X,1,Y) <=> Y=X.
	mul(X,Y,Z1)  \ mul(X,Y,Z2) <=> Z1=Z2.
	mul(X,Y,Z)   ==> delay(X*Y,Z).
	pow(1,X,Y)  <=> Y=X.
	pow(0,_,Y)  <=> Y=1.
	pow(X,Y,Z1) \ pow(X,Y,Z2) <=> Z1=Z2.
	pow(K,X,Y)   ==> delay(X^K,Y).
	add(0,X,Y) <=> Y=X.
	add(X,0,Y) <=> Y=X.
	add(X,Y,Z1)  \ add(X,Y,Z2) <=> Z1=Z2.
	add(X,Y,Z)   ==> delay(X+Y,Z).
	log(X,Y)     ==> delay(log(X),Y).
	exp(X,Y)     ==> delay(exp(X),Y).

	delay(Expr,Res) :- when(ground(Expr), Res is Expr).
\end{prolog-framed}
The rules for "deriv/3" encode the process by which the presence of a
"deriv(L,X,DX)" constraint, interpreted as a request
for the derivative $\hpdbyd[L]{X}$ to be unified with "DX",
is propagated \emph{forward} through each hyperedge (arithmetic
constraint) to which $X$ is an input, using the appropriate form
of the differentiation chain rule for that constraint. This results
in requests for further derivatives until we reach the node "L" itself
and request $\hpdbyd[L]{L}$, which is identically one.

We must also take care to account for the possibility of multiple forward
paths from a variable $X$ to a derived variable $L$. For example, 
the expression
\begin{equation}
	L = 2X + \log X,
	\eqlab{one}
\end{equation}
is encoded as "mul(2,X,Y), log(X,Z), add(Y,Z,L)", or graphically,
\begin{center}
\includegraphics{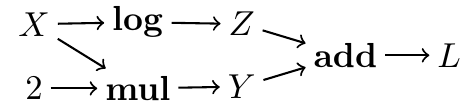}
\end{center}
Each path from "X" to "L" provides an
additive contribution to $\hpdbyd[L]{X}$; for this reason, the following
rules include a mechanism involving constraints "acc/1" (for ``accumulator'')
and "agg/2" (for ``aggregate''), which will be described below.
\begin{prolog-framed}[numbers=left,name=adone,firstnumber=21]
	deriv(L,X,DX) \ deriv(L,X,DX1) <=> DX=DX1.
	deriv(L,L,DL) <=> DL=1.
	deriv(L,_,DX) <=> ground(L) | DX=0.
	deriv(_,_,DX) ==> var(DX) | acc(DX).
	deriv(L,X,DX), pow(K,X,Y) ==> deriv(L,Y,DY), delay(K*X^(K-1),Z), agg(Z,DX).
	deriv(L,X,DX), log(X,Y)   ==> deriv(L,Y,DY), delay(DY/X,Z), agg(Z,DX).
	deriv(L,X,DX), exp(X,Y)   ==> deriv(L,Y,DY), delay(DY*Y,Z), agg(Z,DX).
	deriv(L,X,DX), mul(K,X,Y) ==> deriv(L,Y,DY), delay(DY*K,Z), agg(Z,DX).
	deriv(L,X,DX), mul(X,K,Y) ==> deriv(L,Y,DY), delay(DY*K,Z), agg(Z,DX).
	deriv(L,X,DX), add(X,_,Y) ==> deriv(L,Y,DY),                agg(DY,DX).
	deriv(L,X,DX), add(_,X,Y) ==> deriv(L,Y,DY),                agg(DY,DX).

	acc(X) \ acc(X) <=> true.
\end{prolog-framed}
For example, suppose we request $\hpdbyd[L]{X}$ by inserting "deriv(L,X,DX)"
into a constraint store which already 
contains "mul(2,X,Y), log(X,Z), add(Y,Z,L)".
Using the chain rule, we must include both paths connecting "X" to "L":
\begin{align*}
	\partiald{L}{X} &= \partiald{L}{Y} \partiald{Y}{X} + \partiald{L}{Z} \partiald{Z}{X} 
	= \partiald{L}{Y}\times 2 + \partiald{L}{Z} \frac{1}{X} .
\end{align*}
Hence, the rule for "mul/3" on line 28 inserts "deriv(L,Y,DY)" into the store and
a delayed computation of "DY*2", the result of which, is registered as an additive
contribution to "DX" using "agg/2". The other path is handled by the rule
for "log/2" on line 26, which (renaming the variables to match the example) requests
$\hpdbyd[L]{Z}$ by inserting "deriv(L,Z,DZ)" into the store, along with a delayed computation
of "DZ/X" to provide the other additive contribution to "DX". Let us examine what the constraint
store contains after requesting $\hpdbyd[L]{X}$ (with variables renamed to clarify their
meaning):
\begin{prolog-barred}
	?- mul(2,X,Y), log(X,Z), add(Y,Z,L), deriv(L,X,DX).
	add(Y, Z, L),
	mul(2, X, Y),
	log(X, Z),
	deriv(L, Y, DY),
	deriv(L, Z, DZ),
	deriv(L, X, DX),
	agg(DXFromMul, DX),
	agg(1, DY),
	agg(DXFromLog, DX),
	agg(1, DZ),
	acc(DY), acc(DZ), acc(DX),
	when(ground(X), Y is 2*X),
	when(ground(X), Z is log(X)),
	when(ground(f(Y, Z)), L is Y+Z),
	when(ground(DY), DXFromMul is DY*2),
	when(ground(f(DZ, X)), DXFromLog is DZ/X).
\end{prolog-barred}
As well as delayed goals for evaluating the base expression, and the original
request for $\hpdbyd[L]{X}$, the store also
contains requests for $\hpdbyd[L]{Y}$ and $\hpdbyd[L]{Z}$, three "acc/1" constraints
declaring that "DX", "DY" and "DZ" have additive contributions to be summed,
and four "agg/2" constraints declaring those additive contributions.

The next step is to collect and sum up the additive contributions to each
requested derivative. Aggregation is known to be a slightly awkward process
in CHR \citep{SneyersVan-WeertSchrijvers2007}, and the solution adopted here has a somewhat
imperative flavour:
\begin{prolog-framed}[name=adone,numbers=left,firstnumber=34]
	go \ deriv(_,_,_) <=> true.
	go \ add(_,_,_)   <=> true.
	go \ mul(_,_,_)   <=> true.
	go \ log(_,_)     <=> true.
	go \ exp(_,_)     <=> true.
	go \ pow(_,_,_)   <=> true.
	go \ acc(S)       <=> acc(0,S).
	go <=> true.

	acc(A1,S), agg(T,S) <=> delay(T+A1,A2), acc(A2,S).
	acc(A,S) <=> S=A.
\end{prolog-framed}
Inserting the constraint "go" into the constraint store first removes all the original
arithmetic constraints and all the "deriv/3" constraints so that they do not interfere with
the aggregation process by causing their rules to be fired as various variables are grounded. Then,
each "acc(S)" constraint is replaced with an initial accumulator state represented as
"acc(0,S)". This begins a process by which the rule on line 43 ``mops up'' all the
"agg(T,S)" constraints for a given variable "S", updating the accumulator with a (delayed)
addition of the contribution "T" and the accumulated value so far.
Finally, when no more "agg/2" constraints remain, the target variable "S" is unified with
the value of the accumulator "A" and the "go" constraint removed.
By the end of this process, the constraint store is empty and only frozen goals remain:
\begin{prolog-barred}
	?- mul(2,X,Y), log(X,Z), add(Y,Z,L), deriv(L,X,DX), go.
	when(ground(X), Y is 2*X),
	when(ground(X), Z is log(X)),
	when(ground(X), DXFromLog is 1/X),
	when(ground(f(Y, Z)), L is Y+Z),
	when(ground(DXFromLog), DX is DXFromLog+2).
\end{prolog-barred}
Note how the final expression for "DX" has been partially simplified due to the fact
that the contribution from the "Y" path reduced to the constant 2. Numerical derivatives can
now be computed simply by unifying the input variables (in this case just "X") with numeric
values, for example:
\begin{prolog-barred}
	?- mul(2.0,X,Y), log(X,Z), add(Y,Z,L), deriv(L,X,DX), go, X=2.
	X = 2,
	Y = 4.0,
	Z = 0.6931471805599453,
	L = 4.693147180559945,
	DX = 2.5.
\end{prolog-barred}
If multiple evaluations at different values of "X" are desired, we can either use a backtracking
Prolog aggregator such as "findall/3", or use Prolog's "copy_term/2" to achieve an effect
somewhat like lambda abstraction on the collection of variables and delayed goals. (In both
cases, the frozen goals on the original set of variables remain and are omitted from the
displays below.)
The first approach looks like this:
\begin{prolog-barred}
	?- mul(2.0,X,Y), log(X,Z), add(Y,Z,L), deriv(L,X,DX), go, 
	   time(findall(DX, between(1,1000,X), DXs)).
	DXs = [3.0, 2.5, 2.3333333333333335, 2.25, 2.2, 2.1666666666666665|...]
\end{prolog-barred}
The second requires an auxiliary predicate "copy2/4":
\begin{prolog}
	copy2(X0,Y0,X,Y) :- copy_term(X0-Y0,X-Y).
\end{prolog}
and looks like this:
\begin{prolog-barred}
	?- mul(2.0,X,Y), log(X,Z), add(Y,Z,L), deriv(L,X,DX), go, 
     numlist(1,1000,Xs), time(maplist(copy2(X,DX),Xs,Ds)).
	Xs = [1, 2, 3, 4, 5, 6, 7, 8, 9|...],
	Ds = [3.0, 2.5, 2.3333333333333335, 2.25, 2.2, 2.1666666666666665|...]
\end{prolog-barred}

\section{Observations on the first attempt}

The module "autodiff1" was tested on a differentiation based implementation 
of a generalised inside-outside algorithm, as described in \secrf{intro},
for a probabilistic programming system inspired by PRISM, currently under development.
Performance measurements are deferred until \secrf{performance}, but one minor
observations was that a small but welcome improvement could be obtained by avoiding
conversions from integer to floating point numeric types---to that end, in the
second implementation below, all the integer constants (except for powers in the
"pow/3" constraint) have been replaced with floating point constants.

A more significant shortcoming of the first implementation is that it does not 
support the computation of higher order derivatives, because the numeric
operations required to compute the first order derivatives are implemented at
the lower level of delayed goals, using "delay/2", rather than at the level of the
differentiable constraints "add/3", "mul/3" \etc An attempt to rectify this was
initially successful for the "pow/3" constraint, at least for positive integral
powers, but resulted in nontermination for negative powers and the other
arithmetic constraints.

The reason was that, although the latter \emph{numeric} phase of the computation
propagates \emph{backwards} (triggered by the grounding of variables via 
delayed goals, starting with the value being differentiated and working backwards),
the \emph{analysis} phase of the process propagates \emph{forward}, carried
by the CHR rules for "deriv/3" on lines 25--31. 
A request for $\hpdbyd[L]{X}$ using "deriv(L,X,DX)" generates further requests
for $\hpdbyd[L]{Y}$ for \emph{all} nodes "Y" that "X" contributes to directly in the
computation graph, including any nodes created by
the processing of "deriv(L,X,DX)" itself, even when the forward path through the
new node does not reach the target node "L". For "pow/3" with positive powers, this
recursive process terminates when the power reaches zero, because of the simplification
rule for "pow(0,_,_)" on line 10, but for negative powers and more generally, it does not.
Hence, a rethink was required, which resulted in the method presented in the
next section.

\section{Second attempt: constraint back-propagation}
In the second version, the primary mechanism is the propagation of the "deriv/3" 
constraint backward from the target variable being differentiated, reaching all the 
variables that contribute to the target, regardless of whether or not the derivative 
with respect to that variable was requested; that is, if we want $\hpdbyd[L]{X}$, then
we get $\hpdbyd[L]{Y}$ for all $Y$ that affect $L$. In contrast, the first approach
yields $\hpdbyd[M]{X}$ for all $M$ that $X$ contributes to. The second approach is
more in line with the idea of reverse mode AD.

The code begins, as before, with module and constraint declarations:
\begin{prolog-framed}[name=adtwo]
	:- module(autodiff2, [mul/3, add/3, pow/3, exp/2, log/2, deriv/3, 
	                     back/1, compile/0]).
	:- use_module(library(chr)).
	:- chr_constraint add(?,?,-), mul(?,?,-), log(-,-), exp(-,-), pow(+,-,-),
										deriv(?,-,?), agg(?,-), acc(?,-), acc(-), go, compile.
\end{prolog-framed}
Note that the "go/0" constraint is no longer exported, and instead we have a "back/1"
predicate and a "compile/0" constraint. The idea is that the user indicates 
which derivatives are
required, for example, "deriv(L,X,DX), deriv(L,Y,DY)", as before,
but nothing happens until "back(L)" is called 
to trigger back-propagation starting from "L". Finally, the conversion of 
all the arithmetic constraints,
including those arising from the derivative computations, is triggered by the
insertion of the "compile/0" constraint. This allows the arithmetic simplification rules below to
be fully applied before the lower level arithmetic mechanism is invoked.
\begin{prolog-framed}[name=adtwo,firstnumber=6]
	mul(0.0,_,Y) <=> Y=0.0.
	mul(_,0.0,Y) <=> Y=0.0.
	mul(1.0,X,Y) <=> Y=X.
	mul(X,1.0,Y) <=> Y=X.
	mul(X,Y,Z1) \ mul(X,Y,Z2) <=> Z1=Z2.
	pow(1,X,Y) <=> Y=X.
	pow(0,_,Y) <=> Y=1.
	add(0.0,X,Y) <=> Y=X.
	add(X,0.0,Y) <=> Y=X.
	add(X,Y,Z1) \ add(X,Y,Z2) <=> Z1=Z2.
\end{prolog-framed}
The derivative propagation mechanism is defined below: the user first
indicates that derivatives are required, \eg with "deriv(L,X,DX)".
Since "X" is most likely an input variable to the graph, nothing happens
apart from the insertion of "acc(DX)" into the store.
(If "X" is an intermediate variable,
then back-propagation will occur, but this is harmless.) Then, when
"back(L)" is called, the identity "back(L,L,1.0)" is inserted into the store, 
causing the rules on lines 21--27 to fire progressively backwards through
the graph. The simpagation rule on line 18 ensures any pre-existing "deriv/3"
constraints are absorbed into the process
with the correct unification of the variables representing the derivatives.
\begin{prolog-framed}[name=adtwo,firstnumber=16]
	back(Y) :- var(Y) -> deriv(Y,Y,1.0), go; true.

	deriv(L,X,DX) \ deriv(L,X,DX1) <=> DX=DX1.
	deriv(L,_,DX) <=> ground(L) | DX=0.0.
	deriv(_,_,DX) ==> var(DX) | acc(DX).
	deriv(L,Y,DY), pow(K,X,Y)   ==> deriv(L,X,DX), d_pow(K,X,W), 
	                               mul(DY,W,Z), agg(Z,DX).
	deriv(L,Y,DY), exp(X,Y)     ==> deriv(L,X,DX), mul(Y,DY,T), agg(T,DX).
	deriv(L,Y,DY), log(X,Y)     ==> deriv(L,X,DX), pow(-1,X,RX), 
	                               mul(RX,DY,T), agg(T,DX).
	deriv(L,Y,DY), add(X1,X2,Y) ==> maplist(agg_add(L,DY),[X1,X2]).
	deriv(L,Y,DY), mul(X1,X2,Y) ==> maplist(agg_mul(L,DY),[X1,X2],[X2,X1]).

	agg_add(L,DY,X1)    :- 
		var(X1) -> deriv(L,X1,DX1), agg(DY,DX1); true.
	agg_mul(L,DY,X1,X2) :- 
		var(X1) -> deriv(L,X1,DX1), mul(X2,DY,T1), agg(T1,DX1); true.
	d_pow(K,X,W)   :- 
		K1 is K - 1, KK is float(K), 
		pow(K1,X,XpowK1), mul(KK,XpowK1,W).
\end{prolog-framed}
The aggregation process is defined using "go/0" much as before, and is triggered
automatically by "back/1" after the derivatives have been propagated.
Note, however, that the all arithmetic constraints, those originally inserted
by the user and those generated by the derivative computation, are left
intact in the store, allowing further derivatives to be requested used "deriv/3"
and "back/1".
\begin{prolog-framed}[name=adtwo,firstnumber=36]
	acc(X) \ acc(X) <=> true.
	acc(S1,X), agg(Z,X) <=> add(Z,S1,S2), acc(S2,X).
	acc(S,X) <=> S=X.

	go \ deriv(_,_,_) <=> true.
	go \ acc(DX) <=> acc(0.0,DX).
	go <=> true.
\end{prolog-framed}
Let us examine how the system behaves so far, using the same
example as previously:
\begin{prolog-barred}
	?- mul(2,X,Y), log(X,Z), add(Y,Z,L), deriv(L,X,DX), back(L).
	add(2, DZbyX, DX),
	add(Y, Z, L),
	mul(2, X, Y),
	log(X, Z),
	pow(-1, X, DZbyX).
\end{prolog-barred}
The resulting computation is expressed using the high-level constraints
and can therefore be differentiated further if desired. Graphically, it 
looks like this:
\begin{center}
\includegraphics{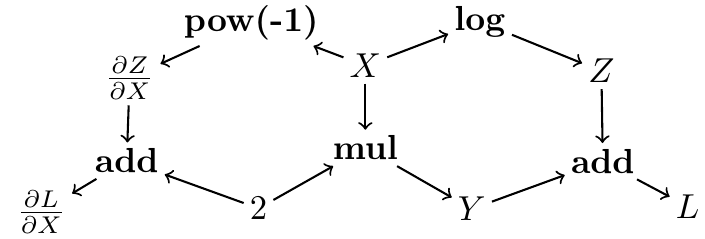}
\end{center}
Note that as drawn here, the data flow is no longer left-to-right, but that this
allows a rather pleasing lateral symmetry to be revealed between $Z$ and $L$ on one side and $\hpdbyd[Z]{X}$
and $\hpdbyd[L]{X}$ on the other. Note also that the $-1$ parameter has been ``bound'' to the 
"pow" operator; this is a reasonable notation since the power must be a ground constant 
in this implementation.

The module is completed with the addition of rules for the "compile/0" constraint, which
replaces all the arithmetic constraints with delayed goals:
\begin{prolog-framed}[name=adtwo,firstnumber=43]
	compile \ add(X,Y,Z) <=> delay(X+Y,Z).
	compile \ mul(X,Y,Z) <=> delay(X*Y,Z).
	compile \ log(X,Y)   <=> delay(log(X),Y).
	compile \ exp(X,Y)   <=> delay(exp(X),Y).
	compile \ pow(K,X,Y) <=> delay(X^K,Y).
	compile <=> true.

	delay(Expr,Res) :- when(ground(Expr), Res is Expr).
\end{prolog-framed}

\section{Example: Taylor series coefficients}
The Taylor series expansion of a function $f:\reals\rightarrow\reals$ 
around a given point $a$ is defined in terms of derivatives
of $f$: if $y=f(x)$, then $f^\prime \equiv \hdbyd[y]{x}$,
$f^{\prime\prime} \equiv \dd^2y/\dd x^2$, \etc, and the Taylor series
representation is
\begin{equation}
	f(x) \approx f(a) + f^\prime(a)(x-a) + \frac{f^{\prime\prime}(a)}{2!}(x-a)^2 
										+ \frac{f^{\prime\prime\prime}(a)}{3!}(x-a)^3 + \ldots
\end{equation}
The coefficients of the powers of $(x-a)$ in this series can be found
using "autodiff2": the sequence of
higher-order derivatives is easily obtained using SWI Prolog's "foldl/4"
high-order predicate to iteratively apply "deriv/3" (via the wrapper "dbyd/4")
to its own output.
After evaluating the derivatives at "X=A", "foldl/6" and "nth_coeff/5" are
used to divide the $k$th coefficient by $k!$.
\begin{prolog-framed}[name=taylor,numbers=none]
	:- use_module(autodiff2).

	derivs(Y,X,[Y|Ds]) :- foldl(dbyd(X),Ds,Y,_).
	dbyd(X,D2,D1,D2) :- deriv(D1,X,D2), back(D1).

	taylor(N,A,X,Y,Cs) :-
	   length(Ds,N), 
	   derivs(Y,X,Ds),
	   compile, X=A,
	   numlist(1,N,Ks),
	   foldl(nth_coeff,Ks,Ds,Cs,1.0,_). 

	nth_coeff(K,D,C,P1,P2) :- P2 is P1*K, C is D/P1.
\end{prolog-framed}
We can test this on some simple functions 
by loading it into SWI Prolog's top level.
For example, if we let $f(x) = 1/(1+x)$, it is easy to verify by
manual differentiation that the coefficients of the Taylor series
around $x=0$ are $[1, -1, 1, -1, \ldots]$. 
Using "taylor/5" to find the first 8 we get:
\begin{prolog-barred}
	?- add(1,X,X1), pow(-1,X1,Y), time(taylor(8,0.0,X,Y,Cs)).
	X = 0.0,
	X1 = Y, Y = 1.0,
	Cs = [1.0, -1.0, 1.0, -1.0, 1.0, -1.0, 1.0, -1.0].
\end{prolog-barred}
We can try a similar a similar test for $f(x) = \log x$, this
time expanding around $x=1$ and requesting twice as many coefficients:
\begin{prolog-barred}
	?- log(X,Y), time(taylor(16,1.0,X,Y,Cs)).
	X = 1.0,
	Y = 0.0,
	Cs = [0.0, 1.0, -0.5, 0.3333, -0.25, 0.2, -0.1667, 0.1429, -0.125|...].
\end{prolog-barred}
Here, the correctness of the result is slightly obscured by the
floating point representation of the result (formatted with limited precision above), 
but it is not difficult to 
modify the program to work with SWI Prolog's rational
number type and obtain the exact answer for this problem, namely the
hyperbolic sequence $[0, 1, -\nf(1,2), \nf(1,3), -\nf(1,4), \nf(1,5), \ldots]$.

More systematic testing with the $f(x)=1/(1+x)$ problem shows that the running
time is quadratic in the number of coefficients requested, 
reaching 0.5 seconds for an expansion of 80 coefficients.

\section{Performance: inside-outside algorithm}
\seclab{performance}

For a medium-scale test of the system, it was used to replace a
direct implementation of the outside algorithm as part of a 
system for working with probabilistic grammars, implemented in Prolog.
A dataset of 30 sentences was sampled from a probabilistic context
free grammar for a small fragment of English, resulting in an inside algorithm
computation consisting of about 1100 multiplications and 400 additions, and
taking 62 parameters as input.

\begin{table}
	\begin{center}
		\begin{tabular}{@{}l@{\quad}llllll@{}}
			\hline \rule{0pt}{1.15em}
						& prolog & autodiff1 & autodiff2 & theano1 & theano2 & tensorflow \\[0.15em]
			\hline\rule{0pt}{1.15em}%
			\emph{Setup} & 0.043  & 0.347     & 0.323     & 9.90    & 1340       & 27.7  \\%
			\emph{Eval}  & 0.0146 & 0.0128   & 0.0089     & 0.116   & 0.0066     & 8.25 \\[0.15em]%
			\hline 
		\end{tabular}
	\end{center}
	\caption{Run time in seconds for the setup and evaluation phases of
	the inside-outside algorithm implemented in six different ways. The table
	reports the shortest run time for each method observed in 5 test runs.}
	\tablab{perf}
\end{table}

Fitting the grammar's parameters to the dataset using an EM algorithm
involves three phases: (i) \emph{setup}---running the inside-outside algorithm over the
data structure resulting from parsing the dataset, but using Prolog variables
for the parameters and delayed goals to defer numerical computations,
(ii) \emph{evaluation}---evaluating the delayed goals 
at the current value of the parameters 
(using the "copy_term" method described in \secrf{autodiff1} to preserve
the delayed goals for multiple use without backtracking),
and (iii) \emph{update}---using the result of the evaluation to update
the parameters. Phases (ii) and (iii) are iterated until convergence.
The performance of phases (i) and (ii) was compared for several implementation
strategies:
\begin{description}
	\item[prolog]
	Inside algorithm in Prolog using delayed arithmetic goals
	and direct implementation of outside algorithm, also with delayed goals.

	\item[autodiff1]
	Inside algorithm in Prolog using "autodiff1" to handle
	the numeric operations and the outside computation by differentiation.

	\item[autodiff2]
	Inside algorithm in Prolog using "autodiff2" to handle
	the numeric operations and the outside computation by differentiation.

	\item[theano1]
		Theano used to build and differentiate inside computation in `fast compile' mode.

	\item[theano2]
		Theano used to build and differentiate inside computation in `fast run' mode,
		which enables optimisation of the computation graph during the setup phase.

	\item[tensorflow]
		TensorFlow (Python interface) used to build and differentiate inside computation.
\end{description}
The experiments were run on a 2012 MacBook Pro with a 2.5GHz Intel Core 
i5 processor and 8GB of memory, and the results are shown in \tabrf{perf}.
Compared with "prolog", both CHR based AD methods 
incur a significant penalty during the setup phase (by a factor of 8 or 
so), with "autodiff2" performing better than "autodiff1", but, as well
as relieving the programmer of the obligation to write an outside algorithm,
"autodiff2" is about 30\% faster than "prolog" in the evaluation phase, probably
because of the algebraic simplifications performed during the constraint 
processing phase. All three Prolog-based implementations were much faster
than either Theano or TensorFlow in the setup phase. Only Theano in
`fast run' mode achieved similar performance in the evaluation phase, but
this was at the expense of a 20 minute setup time. Surprisingly, TensorFlow
took about 8 seconds per evaluation---it is unclear whether or not this
can be put down to the overhead of passing the 62 scalar-valued parameters 
into the TensorFlow session, and receiving the 63 scalar-valued returns.

It should be emphasised, that Theano and TensorFlow were not
designed with this kind of problem in mind: they are intended to be used
with smaller graphs where the nodes represent large multidimensional 
arrays connected by highly parallelisable numerical operations such as
matrix multiplications.

\section{Related work}

Symbolic differentiation has a long history in computer science,
going back to the 60s, with Schoonschip \citep{Veltman1967}, which was
written in assembler and whose code is available on the internet, and
FORMAC \citep{SammetBond1964}, which was written as an extension of FORTRAN.

Prolog was invented by Alain Colmerauer and his group in 1972, and being 
well suited to symbolic computation, 
was applied to the problem soon after, for example, Warren included a short 
program call DERIV in his PhD thesis \citep{Warren1978} and compared it with the 
closest equivalent in LISP.

As a well established field, automatic differentiation has amassed a large body
of literature which is beyond the scope of this note. I will instead focus on
recent developments in machine learning.
Automatic differentiation has received a lot attention in this field---see
\cite{BaydinPearlmutterRadul2015} for a review---
and it forms a core component of modern machine learning frameworks like 
Theano \citep{ChenLiLi2015}, Torch \citep{CollobertKavukcuogluFarabet2011}, 
TensorFlow \citep{MartinAbadiAshishAgarwalPaulBarham2015} and MXNet \citep{ChenLiLi2015}. 
These frameworks are written in languages not very well suited to symbolic manipulation 
and so their code for handling reverse mode AD is quite verbose\footnote{
See, \eg, \url{https://github.com/Theano/Theano/blob/master/theano/gradient.py},
\url{https://github.com/tensorflow/tensorflow/blob/master/tensorflow/python/ops/gradients_impl.py}.
Standards for verbosity in other languages seem to be quite different:
``MXNet is lightweight, e.g. the prediction codes fit into a single 50K lines C++ source file''
\citep{ChenLiLi2015}}.

The language Julia \citep{BezansonEdelmanKarpinski2017} is an interesting 
development in the field of numeric computing, because as well as being designed 
for high performance, it also has a meta-programming facility, with a symbolic data type and
tools for accessing and manipulating the abstract syntax trees of Julia expressions,
which permit the writing of macros not unlike those in LISP.
High performance implementations of AD are available in the packages ForwardDiff
and ReverseDiff\footnote{\url{https://github.com/JuliaDiff}}, while 
Flux\footnote{https://github.com/MikeInnes/Flux.jl} uses macros to transform Julia
code into dataflow graphs for TensorFlow or MXNet. Preliminary tests with the
inside-outside algorithm indicate that ReverseDiff is an order of magnitude faster
than "autodiff2".

\section{Conclusions}

In this technical note, I have described a novel approach to automatic differentiation
using the methods of logic programming. CHR (constraint handling rules) is a very
high level language that enables the logic of reverse mode automatic differentiation to be 
expressed in an extremely concise form. When tested on the problem of parameter learning
in probabilistic context free grammars using the inside-outside algorithm, which
results in a graph of many scalar-valued nodes, the CHR-based implementations
performed far better than the AD tools included with two modern machine learning
frameworks, Theano and TensorFlow. Although not tested against other AD libraries,
this does suggest that the logic programming approach has the potential to be useful
in practical problems and merits further investigation.

	\bigskip
	\noindent
	\textbf{Acknowledgments}\\
	\TheAcknowledgments

	\bibliographystyle{abbrvnat}
	{\bibliography{all,compsci,me}} 
\end{document}